\documentstyle[psfig]{caosp}

\begin{document}
\pubyear{1998}
\volume{27}
\firstpage{441}

\hauthor{I.Kh. Iliev and I.S. Barzova}
\title{Shell signs in the hydrogen-line spectrum of
       some $\lambda$~~Bootis--type stars}
\author{ I. Kh. Iliev \and I. S. Barzova }
\institute{Institute of Astronomy, National Astronomical Observatory,\\
           P.O.B. 136 Smolyan BG-4700, Bulgaria}
\maketitle

\begin{abstract}
The hydrogen-line spectrum of eight $\lambda$ Boo type stars is studied. 
The observed H$_\delta$ and H$_\gamma$ profiles are compared with Kurucz's 
theoretical profiles. The existence of weak emission-like details in the 
cores of five $\lambda$ Boo stars is demonstrated. The Inglis--Teller 
formula is used to calculate the electron densities. It is found that
electron densities in the atmospheres of stars with peculiar H-line
profiles are twice lower than in stars with normal profiles. 
The conclusion is made that stars with peculiar profiles exhibit some of 
the characteristics usually observed in stars with extended atmospheres.  
\end{abstract}

\section{Introduction}

According to Gray (1988), $\lambda$ Boo type stars can be divided into
two groups. Stars in the first group have normal hydrogen-line (NHL) 
profiles, typical of the A dwarfs. The second group contains stars with
peculiar hydrogen-line profiles (PHL). They have weak cores and broad, but
shallow wings. St\"urenburg (1993) has found a significant correlation 
between Gray's classification and his gas shell index. This index is 
related to the presence of sharp circumstellar absorption details in the 
cores of some metallic lines like Ca K and Na D. The PHL stars have strong 
signs of a gas shell, while the NHL stars show only weak indications. 
Thus, the observational data clearly point to a connection between 
the $\lambda$ Boo phenomenon and circumstellar gas. Here we report on the 
effective temperatures and gravities obtained from fits to theoretical 
hydrogen-line profiles, as well as on the gas shell signs found directly 
in the hydrogen-line spectrum of some PHL $\lambda$ Boo stars.

\section{Input data}

Co-added photographic spectra obtained in the range 
$\lambda\lambda$~3600--4800 \AA\AA\ with a moderate resolution 
$(\Delta\lambda/\lambda \sim 20\,000)$ are used for studying the 
hydrogen-line spectrum of eight $\lambda$ Boo stars. Three of them: 
HD\,31295, HD\,125162, and HD\,183324 exhibit NHL profiles, while the others: 
HD 105058, HD 111786, HD 142703, HD 192640 and HD 221756 show PHL 
profiles. 

\section{Hydrogen-line spectrum}

Observed H$_\delta$ and H$_\gamma$ profiles were compared with Kurucz' 
(1993) theoretical profiles computed with realistic (low metal) 
abundances. The line fitting procedure included a bicubic spline 
interpolation scheme and a $\chi^2$--minimization in the two-dimensional 
parametric grid limited by $T_{\rm eff}=$6\,000\,K--10\,000\,K
and $\log g=$3.0--5.0. The fitting procedure did not take in the innermost 
$\pm 3$~\AA\ from the line center of the observed hydrogen lines. 

Effective temperatures and gravities obtained from our `best-fit' 
theoretical profiles are listed in the Table 1. The dashes reflect the fact
that hydrogen-line profiles are not sensitive to gravity below about 
8\,300\,K. Already published data are given for comparison. There are no 
significant differences between the effective temperatures obtained by 
different methods.

\vspace{-4mm}
\begin{table}[htbp]
\scriptsize
\begin{center}
\caption[]{$T_{\rm eff}$ and $\log g$ determined from fitting the H$_\gamma$ 
and H$_\delta$ profiles, compared with the data compiled from 
 different sources. The last two columns contain the numbers of the last
 resolved Balmer line and the corresponding electron densities}
\begin{tabular}{rccccclcc}
\noalign{\smallskip}
\hline\hline
\noalign{\smallskip}
 &&\multicolumn{2}{c} {This study} &
 \multicolumn{2}{c}{Other sources}&& \\ \cline{3-7}
 HD~~&C&$T_{\rm eff}$ & $\log g$ & $T_{\rm eff}$ & $\log g$ & S &
 $n_{\rm max}$ & $\log N_{\rm e}$\\
\hline
\noalign{\smallskip}
 31295&NHL&8900 & 4.2 & 8900 & 4.2 & 1 &18.7& 13.16\\
      &   &  &     & 8550 & 4.0 & 2 &    &      \\
      &   &  &     & 8900 & 4.2 &2$^{+}$&&      \\
      &   &  &     & 8750 & 4.0 & 3 &    &      \\
      &   &  &     & 9000 & 4.2 & 5 &    &      \\
105058&PHL&7900 & --  & 7800 & 3.6 &1$^{+}$&19.5& 13.05\\
111786&PHL& 7700 & --  & 7500 & 3.9 & 1 &20.1& 12.93\\
      &   &  &     & 7600 & 4.0 & 5 &    &      \\
125162&NHL&8600 & 4.1 & 8800 & 4.2 & 1 &18.2& 13.25\\
      &   &  &     & 8400 & 4.0 & 2 &    &      \\
      &   &  &     & 8500 & 4.3 &2$^{+}$&&      \\
      &   &  &     & 8800 & 4.0 & 3 &    &      \\
      &   &  &     & 8900 & 4.1 & 5 &    &      \\
142703&PHL&7500 & --  & 7200 & 4.0 & 1 &19.4& 13.04\\
      &   &  &     & 7200 & 3.9 & 4 &    &      \\
      &   &  &     & 7400 & 4.1 & 5 &    &      \\
183324&NHL&9100 & 4.1 & 9300 & 4.2 & 1 &18.3& 13.23\\
      &   &  &     & 9250 & 4.1 & 5 &    &      \\
192640&PHL&8100 & --  & 8000 & 4.0 & 1 &19.9& 12.96\\
      &   &  &     & 8000 & 3.9 & 2 &    &      \\
      &   &  &     & 8000 & --  &2$^{+}$&&      \\
      &   &  &     & 8150 & 4.0 & 3 &    &      \\
      &   &  &     & 8000 & 4.1 & 5 &    &      \\
221756&(PHL)&8900 & 4.0 & 9000 & 4.0 & 1 &19.7& 12.99\\
      &   &  &     & 9100 & 3.9 & 5 &    &      \\
\noalign{\smallskip}
\hline\hline
\noalign{\smallskip}
\multicolumn{8}{l}{$1$ - uvby$\beta$ (Moon \& Dworetsky (1985)), $1^{+}$ -
uvby$\beta$ with 'unknown' $\beta$ }\\
\multicolumn{8}{l}{$2$ - UBV (Baschek \& Searle, 1969), $2^{+}$ - H$\gamma$
(Baschek \& Searle, 1969) + this study}\\
\multicolumn{8}{l}{$3$ - uvby with fixed $\log g = 4.0$ (Baschek \&
Slettebak, 1988) } \\
\multicolumn{8}{l}{$4$ - Geneva photometry (Paunzen \& Weiss, 1994)}\\
\multicolumn{8}{l}{$5$ - uvby$\beta$ and stellar evolutionary models (Iliev
\& Barzova, 1995)} \\
\end{tabular}
\end{center}
\end{table}

\vspace{-2mm}
\subsection{The residuals}

To pay attention to the hydrogen-line cores which were excluded from 
the fitting procedure, all observed H$_\gamma$ profiles were rectified with 
their theoretical `best-fits'. In other words, the theoretical profiles 
were used as {\em continua}. Residuals obtained for each star after the 
normalization are shown in Figure 1. Weak `emission-like' details were 
found in the bottoms of H$_\gamma$ lines of five $\lambda$ Boo stars.
Taking 
into account the central depths of H$_\gamma$ lines, equivalent widths of 
the weak emissions  in Figure 1 can be found between 80 and 250 m\AA. 
These values are less than two percent of the equivalent width of 
H$_\gamma$ lines. We found similar  details with quite smaller amplitudes
(down to the limit of detection) in H$_\delta$ profiles.

\begin{figure}[htbp]
\parbox[b]{12cm}{\psfig{figure=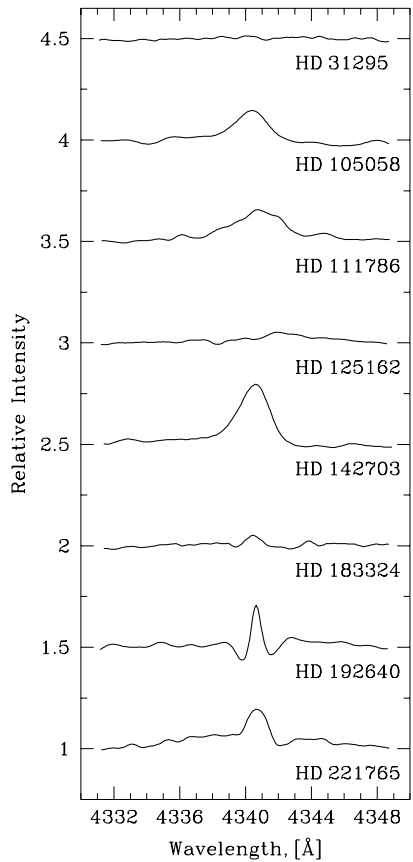,width=5.5cm}}
\hfill
\parbox[b]{5.8cm}
{~~~~~The nature of `emission-like' residuals is not entirely rotational. 
For example, the details found in the H$_\gamma$ lines of HD\,105058 and 
HD\,142703 have nearly the same FWHM, while their rotational velocities 
differ by about 50 percent (130~km\,s$^{-1}$ against 90~km\,s$^{-1}$ 
respectively). The full width of the `emission-like' details at their base is 
larger than the value of $v\sin i$. Similar structures, but on a much larger 
scale, are canonical for the Ae/shell stars. This led us to suggest
that the `emission-like' details could represent a manifestation of gas 
shells or envelopes around the $\lambda$~Boo stars from our list. 
The most interesting case of a shell seen nearly {\em pole-on} in HD\,192640,
which is among the slowest rotating $\lambda$~Boo stars ($v\sin i$ = 35~km\,s$^{-1}$, 
Abt \& Morrell 1995) gives an additional support to this suggestion. Finally, 
all five stars with `emission-like' 
details in their H$_\gamma$ lines are PHL stars. \\
\\
\\}
\caption[]{Isolation of weak emission components in the H$_\gamma$
 lines of five $\lambda$ Boo type stars. The stars that show no peak
 details are NHL stars. Note the shell feature seen nearly pole-on
 in the case of HD\,192640}
\end{figure}

\vspace{-2mm}
\subsection{Electron densities}

We used the Inglis--Teller formula that connects the electron density 
$N_{\rm e}$ in the atmosphere with the number $n_{\rm max}$ of the last 
resolved Balmer line:
\begin{equation}
\log N_{\rm e} = 22.7 - 7.5\log n_{\rm max}
\end{equation}
Both coefficients are taken from Allen (1973). As a rule, the $N_{\rm e}$
value obtained from the Inglis--Teller formula is related to the 
uppermost atmospheric layers where optical depth $\tau \approx 0.1$ and where
the Balmer lines with the highest numbers are formed. In practice, the number 
$n_{\rm max}$ is derived from the relation between Balmer line number 
and central depth $R_{\rm c}$ or equivalent width $W_{\lambda}$. 
An extrapolation of this relation towards the higher numbers gives 
$n_{\rm max}$ as the point where $R_{\rm c}$ (or $W_{\lambda}$)~$\approx 0$ 
(see Fig. 2). It is obvious that in this case $n_{\rm max}$ may be a
fractional number. The values thus obtained for $n_{\rm max}$ have been 
corrected for the rotation and for the spectral class by using the 
coefficients proposed by Kopylov (1961). The error in $n_{\rm max}$ depends 
mainly on the number of points used and on the errors 
in the $R_{\rm c}$ determination. With errors in $R_{\rm c}$ between one 
and two percent the number $n_{\rm max}$ in our case can be determined with 
a standard error of 0.2. This value is consistent with the results of 
Kopylov (1961), who has found the errors in $n_{\rm max}$ between 0.07 
and 0.15. The error of about 0.2 in $n_{\rm max}$ leads to an error in 
the resulting value of $\log N_{\rm e}$ of about 0.05--0.06.

The last visible numbers are listed in Table~1 along with the corresponding 
electron densities. The numbers $n_{\rm max}$ measured by us 
for all the stars studied are in the range 18.2--20.1. The distributions 
of $n_{\rm max}$ suggest that the last resolved 

\begin{figure}[htbp]
\parbox[b]{12cm}{\psfig{figure=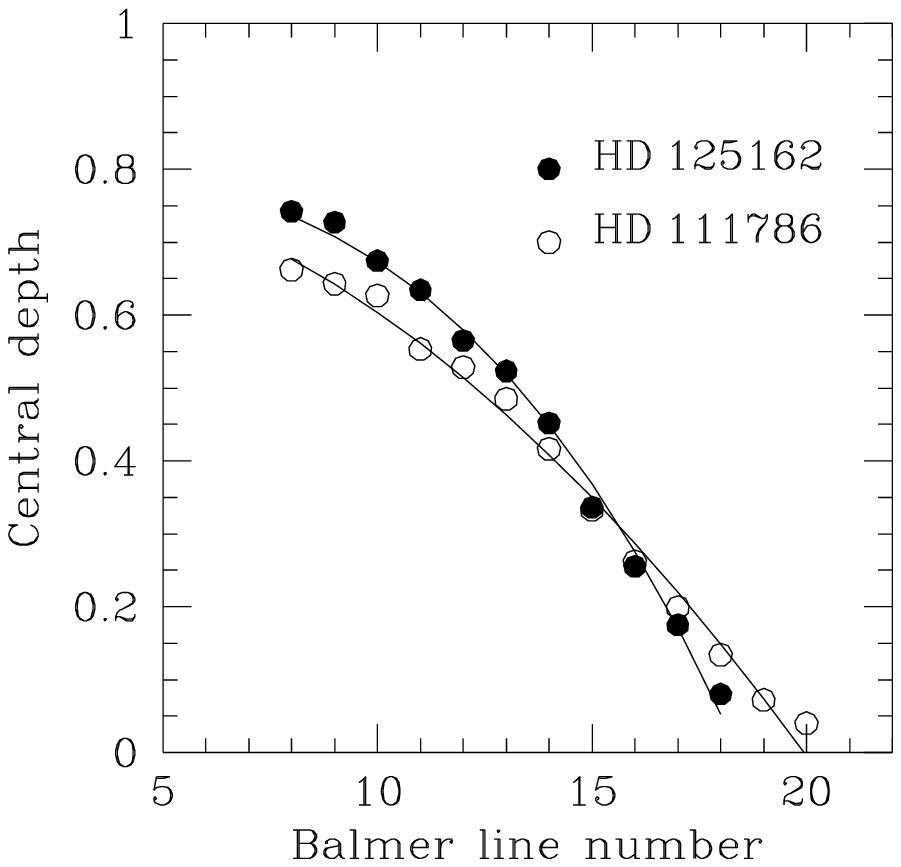,height=5.0cm}}
\hfill
\parbox[b]{6cm}
{Balmer line numbers of
PHL stars are generally larger than in NHL stars.
Non-parametric Wilcoxon 
rank sum test shows that the distributions in the Table 1 are different 
at the 98\% confidence level. The Hodges--Lehmann estimate gives that 
the difference between two medians is 1.1. If this value is interpreted 
in terms of parametric statistics, it will exceed the standard error of 
$n_{\rm max}$ by about 5 times.\\}
\caption[]{The number $n_{\rm max}$ for a given star is determined as the
   point where the interpolated polynomial crosses the abscissa}
\end{figure}

\vspace{-5mm}
\section{Discussion}
\vspace{-2mm}

The main result form the last subsection is that $N_{\rm e}$ 
at the level $\tau \approx 0.1$ in the atmosphere of PHL type stars is 
nearly twice lower than we derived for $\lambda$~Boo stars with 
NHL. This could be interpreted as some evidence that $\lambda$~Boo 
stars with PHL are closer to the end of their evolution {\em on~the~MS}, 
because in the spectral range A0--F0 $n_{\rm max}$ reaches 17.0 for ZAMS, 
18.5 for the middle of the main sequence, and 20.0 for the luminosity 
classes IV--III (Kopylov 1966). An additional support to this 
interpretation could be seen even in the fact that the 
$\log g$ values 
of PHL stars determined from photometric data are in general smaller than 
those of NHL stars. The contradiction with the suggestion of Venn \& 
Lambert (1990) and gas/dust separation scenario developed by Waters et 
al. (1992) (both of them require $\lambda$ Boo stars to be young 
and near the ZAMS) is obvious.

The presence of gas shells or envelopes which surround many $\lambda$ 
Boo stars can change the situation upside-down, making PHL stars 
to bear a resemblance to the well-known Be or Ae/shell stars.
For example, 
Pleione shows a B8V spectrum while in its quiescent mode, while during 
the shell phases $n_{\rm max}$ has reached 40. Thus, the larger 
$n_{\rm max}$ in the spectra of PHL stars can be considered as yet another 
shell symptom. It is important to note that the influence of gas shells on 
the hydrogen-line spectrum of Ae/shell stars which show weak or no emissions 
usually is much more visible in the highest Balmer line numbers. This fact 
could explain why the $\beta$ indices and the corresponding 
effective temperatures of PHL $\lambda$ Boo stars can be close to those obtained 
from spectroscopy, while the Balmer discontinuity, and, therefore, 
luminosity indices like $c_{\rm 1}$, {\em d\ } or $\delta$ can be affected by 
the shell. The conclusion is that $\log g$ values of PHL stars obtained by 
photometry seem to need a revision, because stronger evidences for 
circumstellar shells imply more disturbed $\log g$ values. Unfortunately, 
hydrogen-line profiles offer no alternative solution, since the effective 
temperatures of these stars are mostly below the critical limit of 
about 8\,300\,K.
   
\acknowledgements
This study is supported in part by the National Science Fund under grant
F-603/1996. One of us (I.I.) acknowledges the financial aid from LOC and  
grant VEGA 4175/97.

\end{document}